\documentclass[twocolumn,showpacs,preprintnumbers,amsmath,amssymb]{revtex4}
\usepackage{graphics}
\usepackage{epsfig}
\usepackage{color}
\usepackage{subfigure}

\begin{document}

\title{VUV and X-ray coherent light with tunable polarization from single-pass free-electron lasers}
\author{C. Spezzani$^{1}$, E. Allaria$^{1}$, B. Diviacco$^{1}$, E. Ferrari$^{2,1}$, G. Geloni$^{3}$, E. Karantzoulis$^{1}$, B. Mahieu$^{4,1}$, M. Vento$^{1}$, G. De Ninno$^{4,1}$}
\affiliation{1. Sincrotrone Trieste, S.S. 14 km 163.5, Basovizza (Ts), Italy \\
2. Trieste University, Italy \\
3. European XFEL GmbH, Albert-Einstein-Ring 19, 22761 Hamburg, Germany\\
4. Nova Gorica University, Slovenia\\
}

\date{\today}

\begin{abstract}
Tunable polarization over a wide spectral range is a required feature of light sources employed to investigate the properties of local symmetry in both condensed and low-density matter. Among new-generation sources, free-electron lasers possess a unique combination of very attractive features, as they allow to generate powerful and coherent ultra-short optical pulses in the VUV and X-ray spectral range. However, the question remains open about the possibility to freely vary the light polarization of a free-electron laser, when the latter is operated in the so-called nonlinear harmonic-generation regime. In such configuration, one collects the harmonics of the free-electron laser fundamental emission, gaining access to the shortest possible wavelengths the device can generate. 
In this letter we provide the first experimental characterization of the polarization of the harmonic light produced by a free-electron laser and we demonstrate a method to obtain tunable polarization in the VUV and X-ray spectral range. Experimental results are successfully compared to those obtained using a theoretical model based on the paraxial solution of Maxwell's equations. Our findings can be expected to have a deep impact on the design and realization of experiments requiring full control of light polarization to explore the symmetry properties of matter samples.
\end{abstract}

\maketitle

Variable polarization is a required feature of light sources employed to investigate the properties of matter. The possibility to select light polarization is in particular attractive for those experiments which aim at exploring the local symmetry of the sample under scrutiny, e.g., the lattice geometry of a crystal, the chirality of a molecule, or the presence of a net atomic magnetic moment.
Moreover, several spectroscopic methods rely on the opportunity to choose a well-defined light polarization. Spin \cite{pescia} and angular resolved photoelectron spectroscopy \cite{foto1}, or resonant scattering of polarized x-rays \cite{xray, carra} are some relevant examples.

In all the above-mentioned experiments, light polarization has an influence on the cross section of the electronic transitions one is interested in, when the latter depend on the angular degrees of freedom of the ground and final states \cite{foto2}. Further, in a large class of studies, reversing the helicity of circularly polarized light allows to disentangle the contribution of the signal generated by the breaking of the local symmetry from the one issued by the isotropic background. As an example, one can mention all the experiments based on magnetic circular dichroism \cite{xray}.

For all these reasons, light polarization-dependent spectroscopy using synchrotron radiation has become a powerful tool for studying the electronic and magnetic properties of matter \cite{sut, blume}.

Thanks to a dramatic increase of photon peak-brilliance with respect to conventional sources, as well
as to the possibility of controlling both the temporal duration and the spectral bandwidth of the produced light pulses, free-electron lasers (FEL's) are opening completely new areas in the field of electron spectroscopy \cite{appl1}. Such studies would also greatly benefit from a well-defined and easily-tunable light polarization.

In a FEL, light is emitted by a relativistic electron beam propagating through the static and periodic magnetic field generated by an undulator. As in the case of synchrotron radiation, FEL emission occurs at a given ``fundamental'' wavelength  $\lambda=\lambda_u/(2 \gamma^2)\left(1+K^2\right)$ (where $\lambda_u$ is the undulator period, $\gamma$ stands for the electron-beam relativistic energy and $K$ is the ``deflection parameter'', which is proportional to the strength of the undulator field), and at the harmonics of the latter, i.e. $\lambda_n=\lambda/n$ (where $n>1$ is an integer number). The process leading to FEL emission at $\lambda_n=\lambda/n$ is normally referred to as nonlinear harmonic generation (NHG). 

In fact, for the preparation of the scientific case of these new light sources, a critical question concerns the possibility of generating significant photon flux at FEL harmonics, while maintaining polarization ductility. This would allow to extend towards shorter wavelengths the X-ray spectral region in which FEL's can be used for polarization-dependent experiments. 

In this letter, we demonstrate a method to obtain powerful circularly polarized emission, when the FEL works in NHG configuration. 

\begin{figure}
\centering
{\resizebox{0.48\textwidth}{!}{ \includegraphics{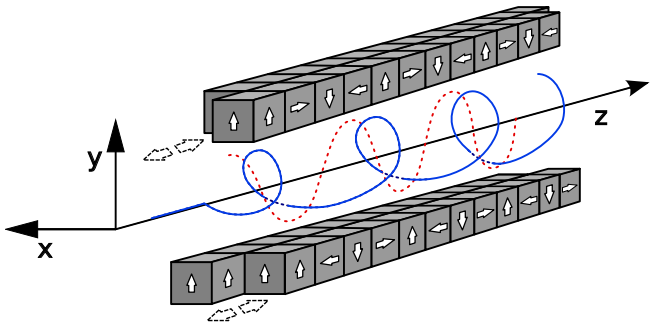}}}
\caption{Schematic layout of a helical (APPLE-type) undulator. When the magnetic arrays are aligned, the magnetic field describes a sinusoid confined in the $(y,z)$ plane (dotted line). The polarization of the emitted light is in this case linear horizontal. Circular polarization is obtained by shifting the magnetic arrays so to make $B_x=B_y$ and the phase between the field components equal to $\lambda_u/4$. In this case, the total magnetic field describes a helix (continuous line).}   
\label{fig1}
\end{figure}

The method relies on the use of a helical undulator \cite{apple}. Contrary to what happens in standard (fixed-polarization) undulators, in helical undulators magnets arrays generating the magnetic field can be shifted with respect to each other, see fig.\ref{fig1}. When the shift is equal to zero, the magnetic field is confined in the vertical $(y,z)$ plane (i.e. $B_x=0$ and $B_y \neq 0$) and describes a sinusoid (see dotted line in fig.\ref{fig1}). Electrons travelling through the undulator (along the $z$ direction) are therefore deflected in the horizontal $(x,z)$ plane and the polarization of the emitted radiation is linear horizontal. This is the so-called planar configuration. By changing the relative position of the magnets' arrays, it is possible to introduce a horizontal component to the magnetic field, $B_x$. When the strength of $B_x$ is equal to that of $B_y$, and the the shift between the two components is equal to $\lambda_u /4$, the resulting magnetic field describes a helix of period $\lambda_u$ (continuous line in fig.\ref{fig1}). In this case, also the electron trajectory follows a helix and the emitted synchrotron radiation is circularly polarized.       
Intermediate values of the ratio $B_x/B_y$ result in an elliptical polarized emission. This kind of undulators is commonly employed in synchrotron storage rings and FELs \cite{nim, fermi} to vary the polarization of the fundamental synchrotron radiation.
As it is well known, in case of planar undulators, on-axis NHG occurs only at odd harmonics, while even harmonics are present only off-axis \cite{Schmitt86}. The NHG was successfully applied as a method to extend the spectral range of FEL's operating with fixed linear polarization, using the third and the fifth harmonics \cite{feldhaus2010}. 

In \cite{geloni, prl1} it has been theoretically and experimentally demonstrated that the NHG signal generated by helical undulators is distributed off-axis. In \cite{nim1}, a quantitative estimate of the off-axis harmonic flux has been performed. Results show that the FEL (coherent) harmonic emission has a quite peculiar angular pattern, compared to the harmonic emission of  (partially coherent) synchrotron radiation. Indeed, while in the latter case off-axis harmonics have a quite significant flux \cite{bruno}, off-axis FEL harmonics are practically vanishing. Such a difference can be explained by considering that, due to the coherent nature of the process, FEL harmonic emission is subject to destructive interference. Coming back to the above-mentioned question about the polarization of FEL harmonics, on the basis of the results reported in \cite{nim1}, one can conclude that collecting off-axis emission does not allow to recover polarization ductility, when the FEL is operated in NHG mode with circularly polarized undulators.

The method we used to obtain polarization ductility in NHG configuration is based on the following consideration: 
changing continuously the ratio between $B_y$ and $B_x$, one can expect to find a satisfactory trade-off between FEL harmonic flux (that goes to zero when approaching circular polarization) and degree of on-axis circular polarization (increasing when moving from linear to circular polarization). Such a method is successfully employed to recover a good degree of circular polarization of on-axis harmonic synchrotron emission \cite{walker, parmi}. However, according to the lesson one can learn from the case of off-axis emission \cite{nim1}, there is no guarantee that such a trade-off exists also in the case of an FEL, i.e. when light emission relies on coherent effects. Indeed, coherence may be expected to have an impact on both the emitted intensity and polarization of harmonic radiation. 

The experiments we present have been carried out using the single-pass FEL installed on the Elettra storage ring \cite{prl2,nim}. The Elettra FEL relies on two independent helical (APPLE-type) undulators (with $\lambda_u = 10$ cm), separated by a dispersive section, and on a Ti:sapphire laser. The NHG emission is generated as follows. The electron beam interacts with the seed laser (or with a harmonic of the latter \cite{hhg}) in the first undulator, called the modulator. The laser-electron interaction results in an energy modulation of the electron beam. The dispersive section is a magnetic device creating a constant (adjustable) magnetic field. When electrons pass through it, the energy modulation generated in the modulator is transformed into spatial modulation, creating electron bunching. The latter is the source of FEL emission, both at the fundamental and harmonic wavelengths. 
Finally, the bunched electron beam is injected into the radiator, where it emits coherent light at the fundamental undulator wavelength and, through NHG, at its harmonics \footnote{In an FEL, the simplest way to generate NHG emission is to inject the electron beam into the radiator immediately after acceleration. Such a method is the standard one when the FEL light is obtained from self amplified spontaneous emission (SASE) \cite{sase1}.  The SASE radiation is characterized by a very good spatial mode, but also by a poor longitudinal coherence. The latter can be drastically improved when the electron beam is pre-bunched as described in the text, prior to injection into the radiator  \cite{yu1, yu2}.}. 

\begin{figure}
\centering
{\resizebox{0.48\textwidth}{!}{ \includegraphics{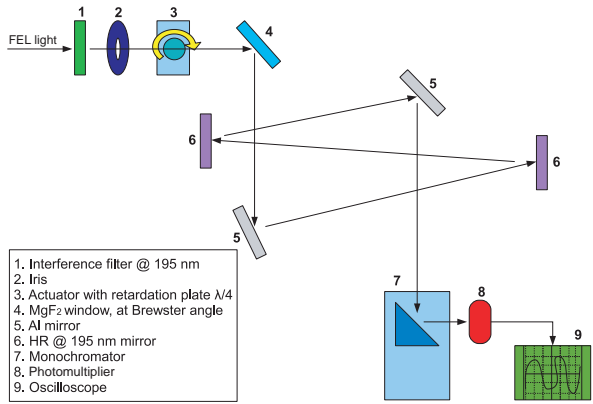}}}
\caption{Scheme of the experimental setup used to measure the polarization of the light emitted by the radiator.}
\label{fig2}
\end{figure}

\begin{figure}
\centering
{\resizebox{0.48\textwidth}{!}{ \includegraphics{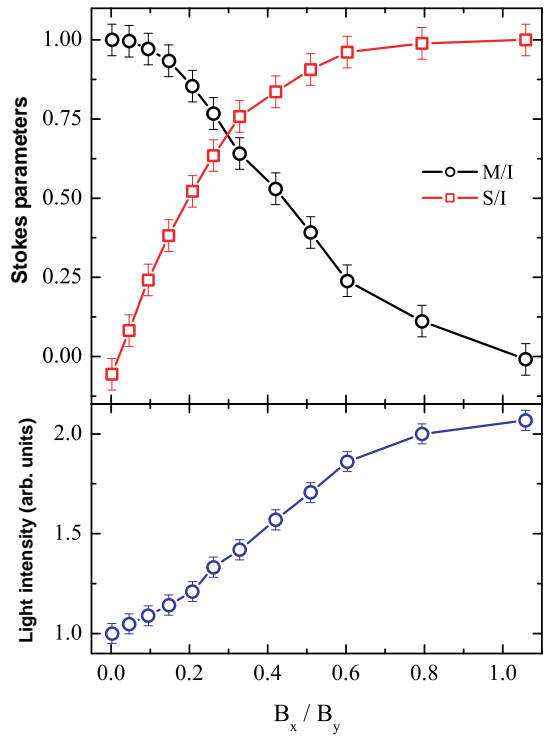}}}
\caption{Fundamental synchrotron emission at 195 nm. Upper panel: normalized Stokes' parameters, $M/I$ (circles) and $S/I$ (squares), as a function of the ratio $B_x$/$B_y$. Lower panel: On-axis intensity (normalized to the intensity recorded when the polarization is linear horizontal), as a function of the ratio $B_x$/$B_y$. Relative errors are estimated to be about 10$\%$.}
\label{fig3}
\end{figure}

The optical setup employed to characterize the polarization of the light emitted by the radiator is shown in fig.\ref{fig2}. The light is first sent through an interferential filter, which strongly attenuates the residual seed radiation. An iris positioned after the filter selects the on-axis emission. Then, the light passes through a quarter-wave plate mounted on a motorized rotation stage, allowing to change the angle $\beta$ of the plate's fast axis with respect to the $y$ axis defined in fig.\ref{fig1} \cite{libro}. After the quarter-wave plate, the light passes through a polarizer that transmits only the light components with linear horizontal polarization. Some other mirrors are then used to further filter out the seed radiation, and to guide the light beam into a monochromator. Finally, the monochromatized signal is acquired by a photomultiplier tube and visualized on a digital oscilloscope. 
Such an optical setup \cite{livi} allows to measure the Stokes parameters \cite{jak} and thus to fully determine the polarization of the light emitted by the radiator. The measurement is performed by acquiring the light intensity transmitted by the setup, $I_T(\beta)$, as a function of $\beta$, for different settings of the FEL's radiator (i.e., different $B_x/B_y$). 
The Stokes parameters $I$, $M$, $C$ and $S$ are found by fitting the measured signal with the following formula \cite{livi}:  
\begin{eqnarray}
I_T(\beta)&&\propto \nonumber \\
&&\frac{1}{2}I -\frac{M}{4} + \frac{1}{2}S\sin2\beta-\frac{1}{4}M\cos4\beta 
-\frac{1}{4}C\sin4\beta\nonumber.
\end{eqnarray}
We remind that $I$ is the total intensity; $M$ represents the fraction of intensity linearly polarized in the coordinate system defined in fig.\ref{fig1}; $C$ is similar to $M$ but in a reference system rotated by 45 degrees; $S$ is the fraction of intensity showing circular polarization. In particular, one finds $M = +I  (-I)$ and $S = 0$ when the polarization is fully linear horizontal (vertical), while $M = 0$ and $S = +I (-I)$ for right hand (left hand) circular polarization. 
In order to test the setup, we performed a preliminary set of measurements of the fundamental synchrotron emission, with the radiator tuned at 195 nm. For all reported measurements, the Elettra storage ring was operated at 900 MeV. 
In Figure \ref{fig3} (upper panel) are plotted the normalized Stokes' parameters, $M/I$ and $S/I$, as a function of the ratio $B_x/B_y$. As expected, when $B_x$ is equal to zero, the polarization of the emitted light is completely linearly (horizontally) polarized. As  $B_x/B_y$ increases, the field component with right-hand circular polarization is enhanced, at the expenses of the linearly polarized fraction. For  $B_x=B_y$, the light is completely right-hand polarized.  
Figure \ref{fig3} (lower panel) shows the on-axis light intensity, as a function of $B_x/B_y$. Like the degree of polarization, also the intensity grows monotonically when passing from linear to circular polarization. This (expected) behavior depends on the fact that the coupling between the electrons and the undulator field, and hence the emission, becomes stronger when passing from linear to circular polarization.
After this set of preliminary measurements, we switched to the characterization of the NHG emission. The FEL modulator was seeded at 390 nm, using the second harmonic of the Ti:Sapphire laser. The optimization of the strength of the FEL dispersive section generated significant bunching at the fundamental wavelength and its harmonics. The radiator was tuned at 585 nm. Since the latter is not a harmonic of the input seed wavelength, no bunching is produced in the modulator at such wavelength. As a consequence, there is no coherent emission at the fundamental wavelength of the radiator. However, one obtains coherent emission through NHG at the second harmonic of the seed wavelength, i.e. 195 nm.
\begin{figure}
\centering
{\resizebox{0.48\textwidth}{!}{ \includegraphics{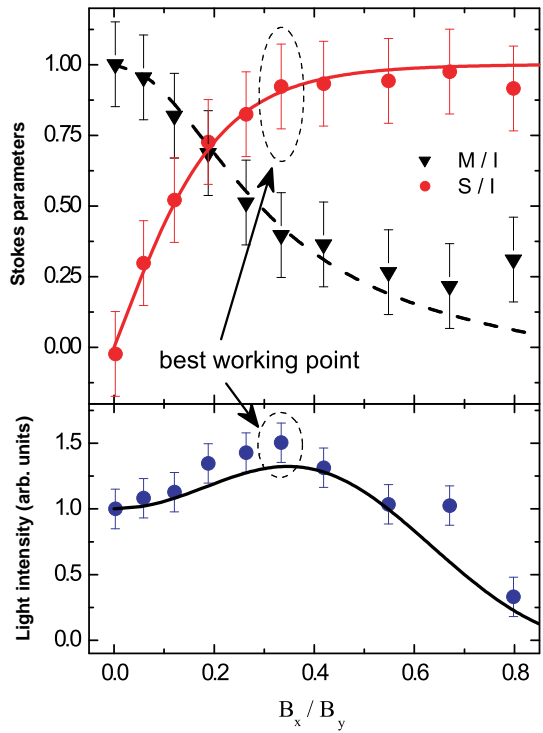}}}
\caption{Nonlinear harmonic emission at 195 nm. Upper panel: normalized Stokes' parameters, $M/I$ (triangles: experiments; dashed line: theory) and $S/I$ (circles: experiments; continuous line: theory), as a function of the ratio $B_x$/$B_y$. Lower panel: On-axis intensity (circles: experiments; continuous line: theory), normalized to the intensity recorded when the polarization is linear horizontal, as a function of the ratio $B_x$/$B_y$. Relative errors are estimated to be about 15$\%$.}
\label{fig4}
\end{figure}
Figure \ref{fig4} reports the same kind of measurements presented in fig.\ref{fig3}, for the case of NHG. The behaviour of the Stokes parameters is similar to the one displayed by the fundamental synchrotron emission: the polarization passes continuously from linear to circular (fig.\ref{fig4}, upper panel) although the curve cannot continue until $B_x=B_y$ because of the sharp drop of the recorded on-axis intensity for $B_x/B_y$ values higher than 0.8 (fig.\ref{fig4}, lower panel). When $B_x=B_y$, corresponding to circular polarization, the on-axis harmonic intensity is zero. The good news is the existence of an intermediate value for $B_x/B_y$, providing a satisfactory trade-off between a relatively high photon flux (about $1.5$ times larger than the flux obtained when the radiator is set for linear polarization) and a relatively high degree of on-axis circular polarization (more than 90$\%$). Polarization purity is enhanced by the natural spatial filtering of the FEL process \cite{nim1}, which suppresses off-axis harmonic emission. In fig.\ref{fig4}, experimental results are compared to those obtained using the theoretical model proposed in \cite{geloni}. In the model, the NHG emission from modulated electrons propagating through a helical undulator is studied by finding the paraxial solution of the corresponding Maxwell's equations. A direct inspection of fig.\ref{fig4} shows that the agreement between theory and experiments is quite satisfactory.
\\
In this Letter we have demonstrated the possibility to generate coherent VUV radiation with fully variable polarization using nonlinear harmonic generation in a free-electron laser based on a helical undulator. The proposed technique, which relies on a suitable tuning of the helical radiator, does not depend on the considered wavelength range, and can be therefore easily extended to X-rays. The result we found is expected to have a deep impact on the scientific case of all new-generation free-electron lasers and, in particular, on the design and realization of the FEL experiments requiring full control of light polarization to explore symmetry properties of matter samples.  
\\
We acknowledge the insightful discussions with F. Parmigiani and W. Fawley, which inspired the idea of the study carried out in this paper. We warmly thank S. Nannarone, A. Giglia and N. Mahne for their advice on the measurement technique.

\end{document}